\documentclass[%
 reprint,
 amsmath,amssymb,
 aps,
pra
]{revtex4-1}

\usepackage{lmodern}
\usepackage{graphicx}
\usepackage{dcolumn}
\usepackage{bm}
\usepackage{hyperref}
\hypersetup{linktocpage,colorlinks,citecolor={blue},pdfdisplaydoctitle=true,pdfpagemode=UseOutlines,bookmarksnumbered=true}
\usepackage{mathrsfs,dsfont}
\usepackage{color}
\usepackage{ upgreek }
\usepackage{slashed}

\newcommand{\bra}[1]{\langle #1 |} 
\newcommand{\ket}[1]{| #1 \rangle } 

\definecolor{cbl}{rgb}{0,0,1}

\definecolor{crd}{rgb}{1,0,0}
 
\newcommand{\upd}{\mathrm{d}}
\newcommand{\tr}{\mathrm{tr}}
\newcommand{\xb}{\mathbf{x}}
\newcommand{\yb}{\mathbf{y}}

\newcommand{\eg}[0]{\textit{e.g.} }

\begin{document}
\title{Comment on ``Spontaneous collapse: A solution to the measurement problem and a source
of the decay in mesonic systems''}

\author{Antoine Tilloy}
\email{antoine.tilloy@mpq.mpg.de}
\affiliation{Max-Planck-Institut f\"ur Quantenoptik, Hans-Kopfermann-Stra{\ss}e 1, 85748 Garching, Germany}
\date{\today}
\begin{abstract}
In a recent article [Phys. Rev. A \textbf{94}, 052128 (2016)], the authors compute the predictions of two collapse models on the transition probabilities of neutral mesons. Notably, they claim to find an influence on the decay rates and attempt to prove that a new parameter $\theta(0)$ is required to fully characterize the noise of standard collapse models. These two claims are incorrect and motivated by flawed computations. This comment derives the correct transition probabilities \emph{exactly} from the master equation, explains how they could be computed perturbatively in a safe way and finally shows where the main mistake of the authors of the original article was made.
\end{abstract}
\maketitle

\section{Introduction}
Spontaneous collapse models may have observable consequences on mesonic systems and studying them in this context is thus a respectable endeavour. In \cite{simonov2016}, the authors attempt to probe the effects of the Quantum Mechanics with Universal Position Localization (QMUPL) model and the Continuous Spontaneous Localization model (CSL) on the oscillation and decay properties of neutral mesons. Their analysis is unfortunately compromised by serious technical errors and conceptual misunderstandings (also present in \cite{simonov2016pla,simonov2017}). The authors find that collapse models modify the decay rates of neutral mesons and that a new parameter $\theta(0)$ is needed to fully characterize the physical content of the models. These two claims are incorrect. Our objective is to redo the computations of \cite{simonov2016} in a safer framework and derive the correct properties of neutral mesons in the presence of a fundamental collapse mechanism.

The authors consider the general continuous collapse equation:
\begin{equation}\label{eq:sse}
\begin{split}
    \upd \ket{\phi_t}= \bigg[-i \hat{H}\,\upd t &+ \sqrt{\lambda} \sum_{i=1}^N\left( \hat{A}_i -\langle \hat{A}_i \rangle_t\right) \upd W_{i,t}\\
    &-\frac{\lambda}{2}\sum_{i=1}^N\left( \hat{A}_i -\langle \hat{A}_i \rangle_t\right)^2\upd t\bigg]\ket{\phi_t},
\end{split}
\end{equation}
where the $\hat{A}_i$ are Hermitian operators and $W_{i,t}$ are independent Wiener processes. This stochastic differential equation (SDE) is understood in the It\^o convention. Once the stochastic integral convention is fixed, equation \eqref{eq:sse} has a unique strong solution and the model is consequently fully specified. There can be no ambiguity in the form of new parameters appearing in quantities computed from \eqref{eq:sse}. This is actually a first hint that there is an issue in \cite{simonov2016}.

The principal objective of \cite{simonov2016} is to compute transition probabilities of the form:
\begin{equation}
P_{\text{in}\rightarrow \text{out}}(t)=\mathds{E}\Big[ \big|\langle\phi_{\text{out}}| \phi_t\rangle\big|^2\Big],
\end{equation}
where $\ket{\phi_0}=\ket{\phi_{\text{in}}}$ and $\mathds{E}[\,\cdot\,]$ denotes the stochastic average. The value of such a probability is, again, unequivocally fixed by the SDE \eqref{eq:sse}. The authors compute such probabilities with a cumbersome and perilous perturbative expansion which leaves an ambiguity (or new degree of freedom) in the results. Such an ambiguity is manifestly spurious as, again, \eqref{eq:sse} entirely fixes the model. Before explaining where the error comes from in sec. \ref{sec:error}, let us first explain how such probabilities can easily be computed with a safe perturbative expansion in sec. \ref{sec:general} and even exactly in sec. \ref{sec:specific} for the collapse models studied in \cite{simonov2016}.

\section{General case}\label{sec:general}
The transition probabilities $P_{\text{in}\rightarrow \text{out}}(t)$ can be computed knowing only the average density matrix $\rho_t=\mathds{E}\big[\ket{\phi_t}\bra{\phi_t} \big]$. At the risk of being overly explicit and repeating well known steps, we detail:
\begin{align}
P_{\text{in}\rightarrow \text{out}}(t)&=\mathds{E}\Big[\tr\big(\ket{\phi_{\text{out}}} \bra{\phi_{\text{out}}}\times \ket{\phi_{t}} \bra{\phi_{t}}\big)\Big]\\
&=\tr\Big(\ket{\phi_{\text{out}}} \bra{\phi_{\text{out}}}\times \mathds{E}\big[\ket{\phi_{t}} \bra{\phi_{t}}\big]\Big)\\
&=\bra{\phi_{\text{out}}} \rho_t \ket{\phi_{\text{out}}}.
\end{align}
It is then well known that $\rho_t$ obeys a linear master equation (ME) of the Lindblad form. The latter is simply obtained by computing $\frac{\upd}{\upd t} \mathds{E}\big[\ket{\phi_t}\bra{\phi_t} \big]$ using \eqref{eq:sse} and It\^o's lemma. The resulting ME reads:
\begin{equation}\label{eq:master}
\frac{\upd}{\upd t}\rho_t= -i[\hat{H},\rho_t] - \frac{\lambda}{2}\sum_{i=1}^N\Big[\hat{A}_i,\big[\hat{A}_i,\rho_t\big]\Big].
\end{equation}
It is a ME encoding decoherence \emph{without} dissipation, as the generators are Hermitian. In many cases of interest, such as the QMUPL and CSL models (in the approximation the authors of \cite{simonov2016} discuss) it can be solved exactly. In the general case, one can find $\rho_t$ and thus transition probabilities perturbatively. For that matter, one goes to the interaction representation to get:
\begin{equation}
\frac{\upd}{\upd t}\rho(t)= - \frac{\lambda}{2}\sum_{i=1}^N\Big[\hat{A}_i(t),\big[\hat{A}_i(t),\rho(t)\big]\Big],
\end{equation}
which is formally integrated in
\begin{equation}\label{eq:interaction}
\begin{split}
\rho(t)= \mathcal{T}&\exp\bigg\{\lambda\sum_{i=1}^N \int_{0}^t \upd s A_i^L(s) A_i^R(s)\\
&- \frac{1}{2}(A_i^L(s) A_i^L(s)+A_i^R(s) A_i^R(s)\bigg\}\cdot \rho(0),
\end{split}
\end{equation}
where $\mathcal{T}$ is the time ordering operator and we have used the standard left-right super-operator notations $A^L_i \cdot \rho = \hat{A}_i \rho$ and $A^R_i \cdot \rho =  \rho\hat{A}_i$. The time-ordered exponential in \eqref{eq:interaction} can be Dyson expanded to compute all the possible transition probabilities as power series in $\lambda$. This approach is applicable to all collapse models and yields unambiguous expansions. One could use it to compute the transition probabilities for the CSL and QMUPL models even without neglecting the Hamiltonian kinetic term as in \cite{simonov2016}.

\section{Collapse models for neutral mesons}\label{sec:specific}
We now focus more specifically on the QMUPL and CSL models with the approximations discussed in \cite{simonov2016}. The authors consider a the simple two level system of a neutral mesons $\ket{M^0}$ and its anti-particle $\ket{\bar{M}^0}$. In the following, we shall neglect the spontaneous decay of these two states into others to simplify the presentation (such a decay can be added by hand in the end anyways). A meson can oscillate between the two aforementioned states and is thus described by a wave-function living in the Hilbert space $\mathscr{H}=L^2(\mathds{R}^3) \otimes \mathds{C}^2$. The Hamiltonian of the system is 
\begin{equation}\label{eq:hamiltonian}
\hat{H}=\mathds{1}\otimes \left(m_L\ket{M_L}\bra{M_L} + m_H\ket{M_H}\bra{M_H}\right),
\end{equation}
where $\ket{M_L}$ and $\ket{M_H}$ are the mass eigenstates taken to be orthogonal:
\begin{align}
\ket{M^0}&=\frac{\ket{M_H} + \ket{M_L}}{\sqrt{2}}\\
\ket{\bar{M}^0}&=\frac{\ket{M_H} - \ket{M_L}}{\sqrt{2}}.
\end{align}
Notice that as in \cite{simonov2016}, the kinetic part of the Hamiltonian is neglected. 

\subsection{QMUPL model}
The QMUPL model is obtained by fixing:
\begin{equation}\label{eq:A}
\hat{A}_i=\hat{q}_i\otimes\left[\frac{m_H}{m_0}\ket{M_H}\bra{M_H} +\frac{m_L}{m_0}\ket{M_L}\bra{M_L}\right]
\end{equation}
where $i$ goes from $1$ to $3$ and $\hat{q}_i$ is the measurement operator for the space coordinate $i$. With this model, the objective of \cite{simonov2016} is to compute the transition probabilities between mass-eigenstates and between particle and anti-particule states. 

The transition probabilities between mass-eigenstates are actually trivial to compute from the very first equation \eqref{eq:sse}. Indeed, it is immediate to see that the stochastic evolution for pure states does not mix different mass eigenstates (and is norm preserving by construction). Equivalently, one sees that the ME \eqref{eq:master} keeps states of the form $\rho = \sigma \otimes \ket{M_{H/L}}\bra{M_{H/L}}$. Hence, one has simply:
\begin{align}
P^{\text{QMUPL}}_{M_{H/L}\rightarrow M_{H/L}}&=1,\\
P^{\text{QMUPL}}_{M_{H/L}\rightarrow M_{L/H}}&=0.
\end{align}
The preturbative results given in equations (16) and (17) of \cite{simonov2016} agree with this straightforward exact computation for $\theta(0)=1/2$. 

To compute the transition probabilities between particle and anti-particle states, we need to solve the ME \eqref{eq:master}. We introduce the following decomposition of $\rho_t$ in position and mass basis:
\begin{equation}
\rho_t=\!\!\!\!\sum_{\mu,\nu = H,L}\int\! \upd^3 \xb\,\upd^3 \yb\,\rho^{\mu,\nu}_t(\xb,\yb) \ket{\xb}\bra{\yb} \otimes \ket{M_\mu}\bra{M_\nu}.
\end{equation}
For the choices \eqref{eq:hamiltonian} and \eqref{eq:A}, the ME \eqref{eq:master} is diagonal:
\begin{equation}
\frac{\upd}{\upd t}\rho^{\mu\nu}_t(\xb,\yb)=\!\!\left[-i(m_\mu\!-m_\nu) \! -\!\lambda \frac{|m_\mu \xb \!-\! m_\nu \yb|^2}{2 m_0^2}\right]\!\rho^{\mu\nu}_t(\xb,\yb)
\end{equation}
and we thus obtain immediately:
\begin{align}
\rho^{\mu\nu}_t(\xb,\yb)=e^{-i(m_\mu\!-m_\nu) \,t -\lambda \frac{|m_\mu \xb - m_\nu \yb|^2}{2 m_0^2} t} \rho^{\mu\nu}_0(\xb,\yb),
\end{align}
where we see that there is manifestly a phase damping term for different mass eigenstates.

The transition probability to a state $\ket{M^0/\bar{M}^0}$ is simply:
\begin{align}
P^{\text{QMUPL}}_{\text{in}\rightarrow M^0/\bar{M}^0}(t)&=\tr\left[ \mathds{1}\otimes\ket{M^0/\bar{M}^0}\bra{M^0/\bar{M}^0} \times \rho_t \right]\\\
&=\frac{1}{2}\int \!\upd^3 \xb \, \rho_t^{HH}(\xb,\xb) +\rho_t^{LL}(\xb,\xb)\nonumber\\
&\hspace{1.3cm}\pm\rho_t^{HL}(\xb,\xb)\pm\rho^{LH}_t(\xb,\xb)
\end{align}
with $\rho_0=\ket{\phi_{\text{in}}}\bra{\phi_{\text{in}}}$. In \cite{simonov2016}, the authors are interested in initial states of the form $\ket{\phi_{\text{in}}} =\ket{\Psi_\alpha}\otimes \ket{M^0}$ where $\ket{\Psi_\alpha}$ is a Gaussian wave-function of width $\sqrt{\alpha}$ in position. For such a state:
\begin{equation}
\int\,\upd^3\xb\,\rho^{\mu\nu}_t(\xb,\xb)= \frac{e^{-i(m_\mu-m_\nu) t}}{2\left(1+\frac{\lambda \alpha(m_\mu-m_\nu)^2}{2 m_0^2}\,t\right)^{3/2}}.
\end{equation}
Hence finally:
\begin{equation}\label{eq:transitionQMUPL}
P^{\text{QMUPL}}_{\alpha,M^0\rightarrow M^0/\bar{M}^0}(t)=\frac{1}{2} \pm\frac{\cos(\Delta m \, t)}{2\left(1+\frac{\lambda \alpha\Delta m^2}{2\,m_0^2}\,t\right)^{3/2}}. 
\end{equation}
Notice again that the total probability is naturally conserved.
This result is exact and the first two terms of the perturbative expansion in $\lambda$ computed in \cite{simonov2016} are found to be \emph{incorrect}, even for $\theta(0)=1/2$ (which, as we shall later argue, is the correct value to put in all the formulas in \cite{simonov2016}). The first two terms in \cite{simonov2016} agree with \eqref{eq:transitionQMUPL} with an exponent $1/2$ instead of $3/2$ presumably because the authors have inaccurately extrapolated from the $1$-dimensional case. Notice that the damping of the oscillations is algebraic, not exponential. However, the details of the long time behavior of the oscillations should be considered with care as the kinetic term of the Hamiltonian is neglected in this approach and the results are heavily sensitive to the initial spreading of the wave function (we are in the ``collapse'' regime described \eg in \cite{bassi2008,bassi2010}).

\subsection{CSL model}
We now derive the correct results for the CSL model. It is obtained by fixing:
\begin{equation}
\begin{split}
    \hat{A}_\xb = \int& \upd^3\yb \,g(\xb-\yb)\, \ket{\yb}\bra{\yb} \\
    &\otimes \left[\frac{m_H}{m_0}\ket{M_H}\bra{M_H} +\frac{m_L}{m_0}\ket{M_L}\bra{M_L}\right]
\end{split}
\end{equation}
where $\xb\in \mathds{R}^3$ is now a continuous index, $g$ is a Gaussian smearing function of width $r_C$ and we have kept the first quantization notations for simplicity. For this model, $\lambda$ is traditionally replaced by $\gamma$.

For the same reason as before, the transition probabilities between mass eigenstates are trivial as the ME does not couple different mass states:
\begin{align}
P^{\text{CSL}}_{M_{H/L}\rightarrow M_{H/L}}&=1,\\
P^{\text{CSL}}_{M_{H/L}\rightarrow M_{L/H}}&=0.
\end{align}

As before, the ME is diagonal in the position / mass eigenstate operator basis:
\begin{equation}
\begin{split}
\frac{\upd}{\upd t}\rho^{\mu\nu}_t(\xb,\yb)=\Big\{&-i(m_\mu\!-m_\nu)  \\
-&\frac{\gamma}{2m_0^2} \Big[(m_\mu^2 + m_\nu^2)\,g*g(0)\\
-& 2\, m_\mu m_\nu \,g*g(\xb-\yb)\big]\Big\}\, \rho^{\mu\nu}_t(\xb,\yb),
\end{split}
\end{equation}
where $g*g$ is the convolution product of $g$ with itself. For a Gaussian, this simply multiplies the variance by $2$ and $g*g(0)=(4\pi r_C^2)^{-3/2}$. As before, we obtain the transition probabilities:
\begin{equation}\label{eq:transitionCSL}
P^{\text{CSL}}_{M^0\rightarrow M^0/\bar{M}^0}(t)=\frac{1}{2} \pm \frac{\cos(\Delta m \, t)}{2} \exp\left[-\frac{\gamma\Delta m^2\, t}{2m_0^2(4\pi r_c)^{3/2}}\right],
\end{equation}
probabilities which do not depend on the spatial profile of the initial wave function. This time, the exact result \eqref{eq:transitionCSL} agrees with the guess (23) of \cite{simonov2016} (again provided $\theta(0)=1/2$). 

\section{Understanding the error}\label{sec:error}
As we have shown, using the ME \eqref{eq:master} is a quick and safe way to compute transition probabilities exactly or perturbatively. Nonetheless, it is important to show what went wrong in the computations made in \cite{simonov2016}. The authors first introduced a new \emph{linear} SDE:
\begin{equation}\label{eq:sselinear}
\begin{split}
    \upd \ket{\phi_t}= \bigg[-i \hat{H}\,\upd t &+ i \sqrt{\lambda} \sum_{i=1}^N\hat{A}_i \upd W_{i,t}
    -\frac{\lambda}{2}\sum_{i=1}^N\hat{A}_i^2\upd t\bigg]\ket{\phi_t}.
\end{split}
\end{equation}
This SDE gives the same ME \eqref{eq:master} as the original SDE \eqref{eq:sse} once averaged at the density matrix level (although, of course, all the ``collapse'' properties are lost). Instead of looking at the ME directly, the authors rewrite \eqref{eq:sselinear} in Stratonovich form:
\begin{equation}\label{eq:strato}
\upd \ket{\phi_t}= \bigg[-i \hat{H}\,\upd t + i \sqrt{\lambda} \sum_{i=1}^N\hat{A}_i \circ \upd W_{i,t} \bigg]\ket{\phi_t},
\end{equation}
where $\int \cdot\,\circ \upd W$ denotes the Stratonovich integral. The authors then attempt to apply perturbation theory on this SDE by replacing stochastic integrals with standard integrals. After astonishingly tedious computations, their final results depend on the integral:
\begin{equation}
I=\int_0^t \mathds{E}\left[\dot{W}_{i,t}\dot{W}_{i,s}\right] \upd s,
\end{equation}
which is ambiguous if one only relies on the rule of thumb that $\mathds{E}\left[\dot{W}_{i,t}\dot{W}_{i,s}\right] \approx \delta(t-s)$ implying $I=1-\theta(0)$. In \cite{simonov2016}, $\theta(0)$ is taken as a new parameter, supposedly further characterizing the noise of the collapse model.

However, as we have previously shown, the model is fully specified by the initial SDE \eqref{eq:sse} and no ambiguity can remain. As we have seen, the correct results (excluding the additional independent error made in \cite{simonov2016} for the QMUPL model) are obtained for $\theta(0)=1/2$. Let us see why this is the case without using the ME. To make sure that perturbation theory is well defined and that we can use standard integrals, we can drive the SDE \eqref{eq:strato} with regularized noise processes $\dot{W}^\varepsilon_{i,t}$:
\begin{equation}
\dot{W}^\varepsilon_{i,t}=\int_{\mathds{R}}\delta_\varepsilon(t-u)\,\upd W_{i,u},
\end{equation}
where $\delta_\varepsilon$ is a smooth function converging to a Dirac delta when $\varepsilon\rightarrow 0$. By virtue of the Wong-Zaka\"i theorem \cite{wong1965}, the solutions of the ordinary differential equation:
\begin{equation}\label{eq:regularized}
\frac{\upd}{\upd t} \ket{\phi_t}=\bigg[-i \hat{H} + i \sqrt{\lambda} \sum_{i=1}^N\hat{A}_i \dot{W}^\varepsilon_{i,t} \bigg]\ket{\phi_t}
\end{equation}
converge to those of the SDE \eqref{eq:strato} \emph{in Stratonovich form} provided the mollifier $\delta_\varepsilon$ is well behaved. A perturbation expansion of $\ket{\phi_t}\bra{\phi_t}$ using \eqref{eq:regularized}, followed by an averaging term by term will then lead to integrals of the form:
\begin{equation}
I^{\varepsilon}=\int_0^t \mathds{E}\left[\dot{W}^\varepsilon_{i,t}\dot{W}^\varepsilon_{i,s}\right] \upd s.
\end{equation}
The latter can be computed exactly with the help of the It\^o isometry $\left(\int f(t) \upd W_t\right)^2=\int f(t)^2 \upd t$ to yield:
\begin{align}
I^\varepsilon&=\int_0^t\upd s\int_\mathds{R} \upd u\,  \delta_\varepsilon(t-u)\delta_\varepsilon(s-u)\\
&=\int_0^t\upd s \int_\mathds{R} \upd u\, \delta_\varepsilon(s+u)\delta_\varepsilon(u)\\
&=\frac{1}{2}\int_{-t}^t\upd s \int_\mathds{R} \upd u\, \delta_\varepsilon(s+u)\delta_\varepsilon(u) \underset{\varepsilon\rightarrow 0}{\longrightarrow} 1/2 .
\end{align}
Consequently, $\theta(0)=1/2$ and it is the only value allowed. Notice that this result requires no symmetry property of the mollifier. Any other choice would correspond to a different underlying SDE.

\section{Conclusion}
Spontaneous collapse models have interesting effects on mesonic systems. They yield decoherence in the mass basis which damps oscillations in the flavor basis. This latter result had been previously preturbatively derived (correctly) in \cite{donaldi2013,bahrami2013}, albeit with the same perilous method as in \cite{simonov2016}. However spontaneous collapse model cannot yield additional decay of neutral mesons beyond what would be put by hand with a phenomenological non-Hermitian Hamiltonian. Further, the SDE \eqref{eq:sse} fully characterizes the model and there is no additional free parameter. We may conclude with a methodological comment. When computing the influence of collapse models on transition probabilities, it is safer to carry perturbative expansions at the master equation level which is a linear equation with regular solutions. Additionally, exact solutions are often easier to find on averaged equations. Finally, if one insists in doing perturbative expansions directly on SDEs (which might be necessary for more complicated theories), smoothing the noise starting from the Stratonovich representation is a good way to lift possible ambiguities.

\bibliographystyle{apsrev4-1}
\bibliography{main}
\end{document}